\documentclass[%
reprint,
onecolumn,
superscriptaddress,
 amsmath,amssymb,
 aps,
prx
]{revtex4-2}

\usepackage{graphicx}
\usepackage{newtxtext}
\usepackage{newtxmath}
\usepackage{color}
\usepackage{natbib}
\usepackage[english]{babel}
\usepackage{bm}
\usepackage[utf8]{inputenc} 
\usepackage{hyperref}
\hypersetup{
    colorlinks = true,
    urlcolor   = blue,
    citecolor  = blue,
}
\usepackage{dsfont}

\newcommand{\Real}{\text{Re}}

\newcommand{\RomanNumeralCaps}[1]

\newcommand{\pd}[2]{\frac{\partial #1}{\partial #2}} 

\newcommand{\bn}{\bm{n}}

\newcommand{\bt}{\bm{t}}
\newcommand{\bu}{\bm{u}}

\newcommand{\de}{\mathrm{d}}
\newcommand{\bee}{\begin{eqnarray}}
\newcommand{\eee}{\end{eqnarray}}
\providecommand{\upi}{\pi}

\begin{document}

\title{Diffusioosmotic corner flows}

\author{Dobromir Nowak}
\affiliation{Faculty of Physics, University of Warsaw, Pasteura 5, 02-093 Warsaw, Poland}
\affiliation{Department of Applied Physics, University of Geneva, Rue de l'Ecole-De-M\'edecine 20, 1205 Geneva, Switzerland}
\author{Maciej Lisicki}
\email{mklis@fuw.edu.pl}
\affiliation{Faculty of Physics, University of Warsaw, Pasteura 5, 02-093 Warsaw, Poland}

\begin{abstract}
{We study flows generated within a two-dimensional corner by the chemical activity of the confining boundaries. Catalytic reactions at the surfaces induce diffusioosmotic motion of the viscous fluid throughout the domain. The presence of chemically active sectors can give rise to steady eddies reminiscent of classical Moffatt vortices, which are mechanically induced in similar confined geometries. In our approach, an exact analytical solution of the diffusion problem in a wedge geometry is derived and coupled to the diffusioosmotic slip-velocity formulation, yielding the stream function of associated Stokes flow. In selected limiting cases, simple closed-form expressions provide clear physical insight into the underlying mechanisms.  Our results open new perspectives for the design of microscale mixing strategies in dead-end pores and cornered microfluidic channels, and offer benchmarks for numerical simulations of confined (diffusio)osmotic systems.}
\end{abstract}

\maketitle

\section{Introduction}

The microscale manipulation of flowing fluids remains at the core of multiple modern applications, from microfluidic appliances in diagnostics and substance testing to small-scale chemical synthesis and industrial precision manufacturing technologies. In Stokes flow conditions appropriate for sub-millimetre fluidic systems, flow can be controlled globally by imposing external forces or pressure gradients that induce laminar flow with only little mixing due to molecular diffusion. While easy to control, global forcing mechanisms pose challenges for applications that require local mixing, selective pumping, or the manipulation of suspended particles within confined environments~\citep{squires2005}. On the other hand, biological and nanorobotic actuation mechanisms, such as ciliated surfaces, rely on localised forcing to achieve swimming, pumping, mixing, nutrient capture, and sensing~\citep{Gilpin2020,Omori2025}. In these settings, surface forcing becomes the key driver of macroscopic bulk flow. In confined geometry, such as in microchannels or pores, this mechanical activity is often coupled to the geometry of the flow domain, and the resulting asymmetry is responsible for creating flow. 

A promising approach exploits phoretic mechanisms, in which surface-generated gradients (of concentration, temperature, etc.) induce effective slip flow on confining surfaces, which in turn gives rise to bulk flow \citep{anderson1989}. Diffusioosmosis and diffusiophoresis, the motion of particles and fluids in response to solute concentration gradients~\citep{golestanian2007,julicher2009,sabass2012}, is now a well-established active propulsion mechanism, with numerous applications in artificial active matter~\citep{Bechinger2016,Shim2022}.

The presence of local chemical gradients in microfluidic channels can lead to cooperation or competition with global advective flow, enabling size-dependent colloid transport~\citep{Shin2015}, or particle focusing that can be precisely tuned through the interplay of channel geometry, confinement, and surface chemical activity~\citep{Ault2018}. {Diffusiophoresis has also been used to organize colloids into sharp bands~\citep{Staffeld1989}, to boost the migration of large particles via imposed solute contrasts~\citep{Abecassis2008}, and to rectify particle motion to yield motility and pattern formation~\citep{Palacci2010}. Subsequent microfluidic studies revealed steady-state focussing in multicomponent gradients~\citep{Shi2016}. In colloids-salt mixtures, phoretic effects were shown to affect mixing~\citep{Raynal2019}, potentially leading to flow effects ranging from enhanced dispersion to blockage in cellular flows~\citep{Volk2022}. The coupling of hydrodynamic flows and phoresis can further tune chemotactic and diffusiophoretic spreading~\citep{Chu2022}.}

In microscale channels, the presence of dead-end pores can be used to induce a concentration difference between the main channel and pores large enough to entrain particles~\citep{Wilson2020}, or capture and retain them~\citep{Battat2019,Akdeniz2023}, affecting filtration and dispersion in {porous media~\citep{Jotkar2024,Doan2021,Chu2021,Sambamoorthy2023,Teng2023,Somasundar2023,Sambamoorthy2025,Alessio2021,Alipour24}}. {Such diffusiophoretic mechanisms have been shown to enhance solute and particle transport into and out of dead-end pores~\citep{Kar2015}, and to enable size-dependent control of colloidal trapping and release via solute gradients in confined geometries~\citep{Shin2015}, with recent simulations and analyses further quantifying phoretic transport and mixing in narrow channels~\citep{Migacz2024,Bhattacharyya2023,Visan2024}.} Active or catalytic pores have also been proposed as local pumps and mixers capable of driving sustained fluid transport without external pressure gradients~\citep{Antunes2022,Antunes2023,Choudhary2025,Migacz2024,Bhattacharyya2023}. When considering catalytic active surfaces, geometric asymmetry alone is sufficient to create heterogeneous concentration fields that can induce propulsion~\citep{michelin2015b,Lisicki2018} or pumping~\citep{michelin2015,Lisicki2016,Michelin2019,Yu2020}. Since recent advances in fabrication allow for precise spatial patterning of active regions on surfaces, for example with catalysts~\citep{Kreienbrink2025,Archer2015}, enzymes~\citep{Sengupta2014}, or surface charges~\citep{Stroock2000,stroock2003}, non-uniform coverage can be used together with geometric features to control microscale flow locally.

Here, we focus on a planar wedge-like geometry of dead-end pores, the walls of which are endowed with chemical activity, and which are filled with viscous fluid. \citet{Moffatt_1964_JFM_18_1,Moffatt_1964_AMS_2_365} was the first to examine the effect of this confinement on a flow that emerges in response to a disturbance that acts far away from the tip of the wedge, as well as from a mechanically active sector on the surface, imposing a slip flow on the boundary. In both cases, the celebrated Moffatt eddies emerge as a solution, with an infinite sequence of vortices being created in the fluid, as later seen experimentally by~\citet{Taneda1979}. Self-similar vortical solutions emerge frequently in externally forced confined flows, such as wedge-shaped trenches with a free surface~\citep{liujoseph1977}, cone-like geometry~\citep{Shankar2005},  electrohydrodynamic flows~\citep{He2022}, simulations of driven cavity flows~\citep{Polychronopoulos2018,Biswas2018}, and in ice flows over subglacial mountain valleys~\citep{Meyer2017}. The classical problem of flow actuation by moving boundaries is perhaps best illustrated in the context of corner flows by the Taylor's scraper problem~\citep{taylor1962}, where {one moving boundary} 'scrapes' the fluid, causing its outward motion along the immobile wall. Moffatt's analysis of corner flow driven by a partial slip on the walls \citep{Moffatt_1964_AMS_2_365}, akin to conveyor belts shearing the fluid locally, is an example of (local) mechanical actuation. Here, we explore a similar concept with a chemical actuation mechanism that couples to flow through {diffusioosmosis}. {We note that the problem of flow in the corner geometry can be treated as a limit of linear elasticity theory for a medium enclosed in wedge-like confinement. For the latter, the Green's functions have recently been found by \citep{DaddiMoussaIder2025_perp,DaddiMoussaIder2025_parallel}. For low-Reynolds-number flows, asymptotic behaviour of the Stokeslet singularity  in a corner were discussed extensively by \citep{Dauparas2018}, and the method of images was used by \citep{Sprenger2023} to explore the dynamics of confined microswimmers.}

In this work, we consider a non-uniform coverage of the wedge with a catalyst that induces the release or capture of solute. The heterogeneous concentration field, emerging from geometric asymmetry of the fluid domain, drives slip flow on the active surfaces, which in turn induces bulk flow that takes the form of a sequence of vortices. We calculate the flow analytically in the simplest cases. Next, using the Mellin transform formalism, we present an approach that applies to any coverage of the walls with chemical activity. 

The paper is structured as follows. First, in Sec. \ref{sec:phoretic-flow}, we present the general mathematical framework of {diffusioosmotic} Stokes flows, describe the geometry of the problem, and the relevant physical quantities. In Sec. \ref{sec:corner-flows}, we discuss the general solutions of the diffusion equation for the solute and the biharmonic equation for the flow stream function in the wedge geometry. We also introduce the formalism of Mellin transforms suited to the geometry considered. Next, in Sec. \ref{sec:uniform-walls} we present analytical solutions for the {diffusioosmotic} flows induced by uniform coverage of one or both walls with a catalyst. In the following Sec. \ref{sec:active-absorbing} we discuss the solute concentration field emerging with a single active sector on one of the walls and the other boundary being absorbing, for different geometric settings. The complementary problem of diffusion with an active sector and a reflective boundary is treated in Sec. \ref{sec:active-reflective}. Finally, we show in Sec. \ref{sec:biharmonic} how the solute concentration fields translate to the flow, and obtain the flow field numerically by solving for the stream function. Additionally, in Sec. \ref{sec:neumann} we discuss the case of the chemical activity of the walls given by analytical functions rather than patches, represented by step-like profiles. We conclude the paper in Sec. \ref{sec:conclusions}.

\section{{Generation of diffusioosmotic flows}}\label{sec:phoretic-flow}

\subsection{General framework}
We adopt a continuum description of {diffusioosmotic} transport, following established theoretical frameworks ~\citep{golestanian2007,julicher2009,sabass2012}, to analyse the two-dimensional flow field generated within a wedge formed by two semi-infinite lines starting from one point. 
The fluid within the wedge is characterised by dynamic viscosity $\eta$, density {$\rho_0$}, and contains solute of local concentration {$\tilde{C}$ (number of particles per unit volume)}, with diffusivity $\kappa$. 
The chemical activity $\mathcal{A}$ of a surface $\mathcal{S}$ quantifies the fixed rate of solute release ($\mathcal{A}>0$) or absorption ($\mathcal{A}<0$) on the surface by
\begin{equation}\label{fixedflux}
\kappa \bn\cdot\nabla {\Tilde{C}} =-\mathcal{A}\quad \textrm{      on   } \mathcal{S},
\end{equation}
{where $\bn$ is the normal unit vector on $\mathcal{S}$.}
Due to the short-range interaction of solute molecules with the cavity boundary, local concentration gradients result in the motion of the solute and, consequently, drive the motion of the fluid~\citep{anderson1989}. Assuming that the thickness of the interaction layer is small compared to the dimensions of the cavity, the classical slip-velocity formulation can be used~\citep{michelin2014}, and local solute gradients induce an effective slip velocity on the boundaries
\begin{equation}\label{slip_def}
{\tilde{\bu}} =\mathcal{M}(\boldsymbol{1}-\bn\bn)\cdot\nabla {\tilde{C}}\quad \textrm{      on   } \mathcal{S},
\end{equation}
which drives the bulk motion of the fluid. Here, 
 $\mathcal{M}$ is the local phoretic mobility at the surface $\mathcal{S}$. It is related to the surface-solute interaction potential \citep{anderson1989}.

Introducing $\mathcal{R}$ as the characteristic length scale, and $\mathcal{U}=|\mathcal{AM}|/\kappa$ as the characteristic phoretic velocity generated along $\mathcal{S}$, we can define the P\'eclet and Reynolds numbers
\begin{equation}
    \mathrm{Pe}=\frac{\mathcal{UR}}{\kappa}, \qquad \text{Re}=\frac{ {\rho_0}\, \mathcal{U} \mathcal{R}}{\eta}.
\end{equation}
which quantify, respectively, the relative importance of solute advection and diffusion as transport mechanisms and effects of inertia and viscosity in forces that shape the flow. When $\text{Pe}$ is small enough, the solute dynamics are purely diffusive and are governed by Laplace's equation
\begin{equation}\label{laplacec}
\nabla^2 {\tilde{C}}= 0,
\end{equation}
in the whole domain. Provided that inertial effects are negligible (i.e. $\text{Re}\ll 1$), the flow and pressure fields satisfy the incompressible Stokes equations:
\begin{align}\label{Stokes}
\eta\nabla^2 {\tilde{\bu}} &=\nabla {\tilde{P}}, \\
\nabla\cdot{\tilde{\bu}} &= 0.
\end{align}
The diffusive Laplace's problem for the solute concentration {$\tilde{C}$} effectively decouples from the hydrodynamic problem and may be solved independently. The concentration {$\tilde{C}$} can be used to compute the slip flow along $\mathcal{S}$ via Eq.~\eqref{slip_def}, which serves as the boundary condition for the flow field in Eq.~\eqref{Stokes} within the cavity. 

{We note here that while the fluid is assumed incompressible, in phoretic problems involving the transport of particles, effective particle velocity fields can appear compressible due to concentration-dependent drift. For instance, \citet{Raynal2018} showed that diffusiophoretic drift can induce an effectively compressible colloid flow, even in an incompressible solvent, leading to transient particle focusing. Similarly, \citet{Chu2020} studied colloid transport under transient solute gradients where spatially varying drift produces apparent compression or expansion, though the underlying fluid remains incompressible.}

Given the boundary conditions that we assume in this problem, and the governing Laplace equation, if $\tilde{C}_0$ is a constant and the concentration field {$\tilde{C}$} is a solution, then {the excess concentration $\tilde{c} = \tilde{C} - \tilde{C_0}$} is a solution too. The boundary conditions~\eqref{fixedflux},~\eqref{slip_def} do not depend on the absolute value of the local concentration. For all {$\tilde{C}_0$}, the resulting flow will be the same. This also implies that {$\tilde{c}$} can be negative. After shifting by a positive {$\tilde{C}_0$}, we thus will still obtain a physical solution. In the next sections, we will therefore {replace concentration $\tilde{C}$ with field $\tilde{c}$.}
\subsection{Nondimensionalization of the model}
We have already defined $\mathcal{U}$ and $\mathcal{R}$ as the characteristic velocity and length. {The choice of the characteristic length depends on the problem at hand––we discuss this in detail when defining specific problems, but in general it may be set e.g. by the size of an active patch on the wall.} {Since $\tilde{c}$ is the excess concentration, its variations scale with the magnitude of the gradients produced at confining boundaries, which are induced by surface chemistry. Its natural scale is therefore ${\mathscr{C}=}|\mathcal{A}|\mathcal{R}/\kappa$, where $|\mathcal{A}|$ is the typical magnitude of the chemical surface activity.} {The characteristic pressure is  constructed from the typical velocity scale, given by the mobility and the magnitude of concentration, and reads ${\mathscr{P}=}\eta |\mathcal{AM}|/\mathcal{R}\kappa$}. {The dimensionless pressure, concentration and velocity fields are thus given by $P={\tilde{P}}/{\mathscr{P}}$, $c={\tilde{c}}/{\mathscr{C}}$, 
and $\bu = {\tilde{\bu}}/{\mathcal{U}}$. Similarly, the dimensionless activity and mobility are given by $A = \mathcal{A}/|\mathcal{A}|$ and $M = \mathcal{M}/|\mathcal{M}|$. Since activity can vary over the surface $\mathcal{S}$, the choice of the characteristic activity is not always obvious. Because $\textrm{Pe},\textrm{Re} \propto |\mathcal{A}|$, choosing the maximal value of $|\mathcal{A}|$ on the surface $\mathcal{S}$ leads to the strongest conditions on $\textrm{Pe}$ and $\textrm{Re}$.
For the diffusion of solute, the nondimensional governing equations are as follows}
\begin{subequations}
\begin{align}
    \label{laplacec2} \nabla^2 c &= 0, \\
    \label{fixedflux_nondim}
\bn\cdot\nabla c &=-A\quad \textrm{      on   } \mathcal{S}, \\
\intertext{while the flow problem becomes}
   \nabla^2 \bu &=\nabla P, \label{eq:Stokes1} \\
   \nabla\cdot\bu &= 0, \label{eq:Stokes2} \\
\bu &=M(\boldsymbol{1}-\bn\bn)\cdot\nabla c\quad \textrm{      on   } \mathcal{S}.
\end{align}
\end{subequations}

In the following, we apply this general framework to the specific geometry of a narrowing corner to explore the ways in which fluid motion can be actuated within such a cavity by purely chemical means, with no moving mechanical parts.

\section{Corner flows} \label{sec:corner-flows}

The problem of {osmotic} flow generation in {a} wedge qualitatively resembles that leading to Moffatt eddies \citep{Moffatt_1964_JFM_18_1,Moffatt_1964_AMS_2_365}. In the classical problem analysed by \citet{Moffatt_1964_JFM_18_1}, an infinite sequence of eddies is formed in a wedge-shaped planar domain filled with viscous fluid due to a disturbance acting at asymptotically large distances from the corner. Eddies emerge as a self-similar solution to a second-kind eigenvalue problem for the stream function. However, similar flow structures can arise when the fluid is actuated by a moving boundary, as in the famous Taylor's scraper problem \citep{taylor1962}, where one of the walls of the wedge-shaped cavity moves with constant speed and drives macroscopic flow. A refined variant of this geometric setting involves motion of a portion of the boundary, when a section of it is endowed with slip velocity, as explored by \citet{Moffatt_1964_AMS_2_365}. Using Mellin transform techniques, Moffatt found a solution of this problem for a single moving region (and two regions placed symmetrically), corresponding to transmission belts mounted within the walls. The solution again involves a sequence of corner eddies. This mechanical example operates using the same principle as one implemented in our case.

Here, we focus on a similar geometry but allow the walls to exhibit chemical activity, which drives the flow. The {diffusioosmotic} flow generation mechanism leads, in general, to a non-uniform slip velocity profile on the bounding walls. However, for the case of uniform coverage of either one or both walls with catalyst, the induced slip velocity remains constant, therefore directly reducing to the previously obtained results. As we demonstrate analytically in the following, both antisymmetric and symmetric flows can be induced, depending on the specific coverage pattern. 

\subsection{Solute concentration}

\begin{figure}
\centering
\includegraphics[width=0.6\textwidth]{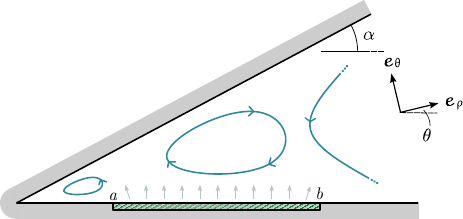}
\caption{Geometry of the {diffusioosmotic} corner flow setup {in polar coordinates $(\rho,\theta)$.} {In a wedge of opening angle $\alpha$}, an active patch on the $\theta=0$ wall covering the radial section $\rho\in[a,b]$, releases solute (grey arrows) and generates an inhomogeneous concentration field which drives circulatory flow indicated by schematic streamlines.  \label{fig:scheme}}
\end{figure}

We first focus on the problem of diffusion of a solute in the corner domain. Upon introducing cylindrical polar coordinates $(\rho,\theta)$ (see figure \ref{fig:scheme}), stationary Laplace's equation for the concentration field $c$ takes the form
\begin{equation}
{\rho}\pd{}{\rho}\left(\rho\pd{c}{\rho}\right) + \pd{^2 c}{\theta^2} = 0. \label{eq:laplace}
\end{equation}
Suppose that the wedge extends in polar coordinates from $\theta=0$ to $\theta=\alpha$.
The boundary conditions are imposed on the normal gradient of concentration, $\bn\cdot\nabla c$, and are thus given by the chemical activity distribution on the walls, $A_0(\rho)$ and $A_\alpha(\rho)$ as
\begin{equation}
	\frac{1}{\rho}\pd{c}{\theta} = -A_0(\rho),
 \label{condition_0}
\end{equation}
if the condition is imposed on the wall $\theta=0$, and
\begin{equation}
	\frac{1}{\rho}\pd{c}{\theta} = A_\alpha(\rho),
 \label{condition_alpha}
\end{equation}
if the condition is imposed on the wall $\theta=\alpha$. {For the 2D diffusion equation to be well-posed, we shall require the integrals of both activity functions over the walls to add to zero, as we discuss later in the manuscript.} 

\subsection{Stokes flow}
Once the concentration distribution $c(\rho,\theta)$ is known, the slip flow velocity $\bu_s = u_s \bm{e}_\rho$ on the boundary is determined as 
\begin{equation}
	u_s = M\left.\pd{c}{\rho}\right|_\mathcal{S}, 
\end{equation}
which becomes the boundary condition for a two-dimensional flow in the fluid domain. In this problem, it is convenient to introduce the {two-dimensional} stream function {$\Psi(\rho,\theta)$~\citep{Batchelor2000,Deville2022_Stokes_flows} such that in polar coordinates}
\begin{equation}
    u_\rho = \frac{1}{\rho}\pd{\Psi}{\theta}, \quad u_\theta = -\pd{\Psi}{\rho}.
\end{equation}
{Since $\bm{u}=(u_\rho,u_\theta)$ satisfies the incompressible Stokes equations~\eqref{eq:Stokes1}-\eqref{eq:Stokes2}, $\Psi$ obeys the biharmonic equation}
\begin{equation}
	\nabla^4 \Psi = 0,
    \label{eq:biharmonic}
\end{equation}
with the boundary conditions on the two planes $\theta = \{0 , \alpha\}$ being
\begin{equation}\label{eq:bc}
	u_\rho = \frac{1}{\rho}\pd{\Psi}{\theta} = u_s, \qquad u_\theta = -\pd{\Psi}{\rho} = 0, \qquad \text{for}\  \theta= \{0 , \alpha\}.
\end{equation}

\subsection{Solution of Laplace and biharmonic equations in the Mellin space}

We shall now introduce the formalism of {Mellin transforms~\citep{Butzer1997_Mellin_transf, Debnath2016_integral_transf}} to solve the more general case of a single active sector. Indeed, Mellin transforms have become the traditional method for solving boundary-value problems in wedge-shaped regions~\citep{TRANTER1948,Moffatt_1964_JFM_18_1,Moffatt_1964_AMS_2_365,Martin2017}. In cylindrical coordinates, Laplace's equation takes the form of Eq. \eqref{eq:laplace}. 
{We denote by $\bar{c}(p,\theta)$ the Mellin transform $\mathscr{M}_p$ of concentration $c(\rho,\theta)$, where $p$ is the transform variable and}
\begin{equation}\mathscr{M}_p\{c(\rho,\theta)\}:=  \int_{0}^{\infty}\rho^{p-1} c(\rho,\theta) \,\de\rho =: \bar{c}(p,\theta). \end{equation}
In addition, $\mathscr{M}^{-1}$ is the inverse Mellin transform, which reads
\begin{equation}
c(\rho,\theta)=\mathscr{M}^{-1}\{\bar{c}(p,\theta)\}=\frac{1}{2\upi i}\int_{\gamma-i\infty}^{\gamma+i\infty} \bar{c}(p,\theta)  \rho^{-p} \de p. \label{eq:reverse_Mellin}
\end{equation}
{The conditions for the existence of the inverse transform are that for a chosen parameter $\gamma$, the integral $\int_{0}^{\infty}\rho^{\gamma-1} c(\rho,\theta) \,d\rho$ exists and the {complex} integration line $(\gamma-i \infty, \gamma+i\infty)$ lies within the strip of analyticity of $\bar{c}(p,\theta)$. When these conditions are satisfied, the inverse is independent of $\gamma$. Unless stated otherwise, we take $\gamma=0$.} Then, for appropriate $c(\rho,\theta)$, we have
\begin{equation}
\mathscr{M}_p\{\rho\partial_\rho c(\rho,\theta)\}=\mathscr{M}_{p+1}\{ \partial_\rho c(\rho,\theta)\}=-p\bar{c}(p,\theta),
\end{equation}
where we introduced an abbreviation for derivatives, $\partial_\rho \equiv \partial/\partial\rho$, etc. The Laplace equation for concentration, 
\eqref{eq:laplace}, implies that $p^2\bar{c}+{\partial^2_\theta} \bar{c}=0$ and for some $C(p)$, $D(p)$ we have
\begin{equation} \bar{c}(p,\theta)= C(p) e^{ip\theta}+ D(p)e^{-ip\theta}.
\label{eq}
\end{equation}

We now turn to the flow problem. Let $\Psi$ be the stream function. The biharmonic equation, $ \nabla^4 \Psi =0$, in cylindrical coordinates and after applying the Mellin transform $\mathscr{M}_{p+4}$ becomes
\begin{equation}
    \left\{\partial_\theta ^4 + [(p+2)^2 +p^2]\partial_\theta^2 + p^2(p+2)^2\right\}\bar{\Psi}(p,\theta)=0,
\end{equation}
where we write $\bar{\Psi}(p,\theta):=\mathscr{M}_{p}\{ \Psi(\rho, \theta)\}$. The general solution then reads
\begin{equation}
    \bar{\Psi}= F_1(p) \cos(p\theta)+ F_2(p)\sin(p\theta)+G_1(p)\cos((p+2)\theta)+ G_2(p)\sin ((p+2)\theta).
    \label{psi_general}
\end{equation}

\section{Phoretic walls with uniform coverage} \label{sec:uniform-walls}

To study the effect of chemical activity on corner flows, we first focus on the case when the coverage by catalyst is uniform, and the heterogeneity of the concentration field stems purely from the geometry of the wedge. In the two examples studied in the following, either one or both walls are active, and the problem admits analytical solutions. {In this case, the problem has no natural length scale, and the velocity scale is given purely by the activity and mobility at the surface.}

\subsection{A single active wall}

In the first simple example, we consider a single active wall, with constant activity $A$ on the surface. To ensure the existence of a steady-state solution, we assume that the other boundary is absorbing, and thus we can write the boundary conditions for the concentration problem as
\begin{align}
\pd{c}{\theta}= -A\rho\qquad  &\text{for $\theta = 0$}, \\
c = 0 \qquad &\text{for $\theta = \alpha$}.
\end{align}
Using the separated form $c=R(\rho)\Theta(\theta)$ as an ansatz, we find the general solution 
\begin{equation}
	R(\rho) = {r_1} \rho^\lambda + {r_2} \rho^{-\lambda}, \qquad \Theta(\theta) = {q_1}\cos\lambda\theta + {q_2} \sin \lambda\theta, 
\end{equation}
{where $r_1$, $r_2$, $q_1$, $q_2$, $\lambda$ are constants which need to be determined.}
Applying the boundary conditions, we obtain the concentration field as
\begin{equation}
	c(\rho,\theta) =- \frac{A \rho \sin(\theta - \alpha)}{\cos\alpha}.
\end{equation}
{The fact that the concentration increases linearly with the distance from the tip is a consequence of the fixed flux boundary condition and the fact that the active patch extends to infinity. In reality, the finite size of the active patch limits the concentration field, as we discuss later on.}
{As a result, the tangential gradient of the concentration profile yields a constant slip velocity distribution on the surface,}
\begin{align}
\bu= MA \tan \alpha\ \bm{e}_\rho\qquad &\text{for $\theta = 0$}, \\
\bu=\bm{0} \qquad  &\text{for $\theta = \alpha$}.
\end{align}
With this velocity profile, the problem becomes identical to the well-known Taylor's scraper \citep{taylor1962}, with the imposed slip velocity $U=MA\tan\alpha$. The stream function then has the form
\begin{equation}
	\Psi = U \rho {F}(\theta),
\end{equation}
where ${F}(\theta)$ satisfies an ordinary differential equation
{
\begin{equation}
	F{''''} + 2 F{''} + F = 0, \label{scraper_ode}
\end{equation}
}
with the boundary conditions {$F(0) = 0$, $F'(0)=1$, $F(\alpha)=0$, $F'(\alpha)=0$.} {We note that by Eq.~\eqref{eq:bc} the boundary conditions for {$F$} correspond to the normal velocity at the surface, while those for {$F'$} pertain to the induced slip (tangential) velocity component}. This is a particular form of the general solution to planar elasticity problems for the biharmonic Airy stress function due to \cite{michell1899}. The resulting stream function reads:
\begin{equation}
\Psi = U \rho {F}(\theta) = \frac{U \rho}{\alpha^2 - \sin^2 \alpha}\left[ \alpha (\alpha-\theta)\sin\theta-\theta \sin\alpha \sin(\alpha-\theta) \right].
\end{equation}

\begin{figure}
\centering
\includegraphics[width=\textwidth]{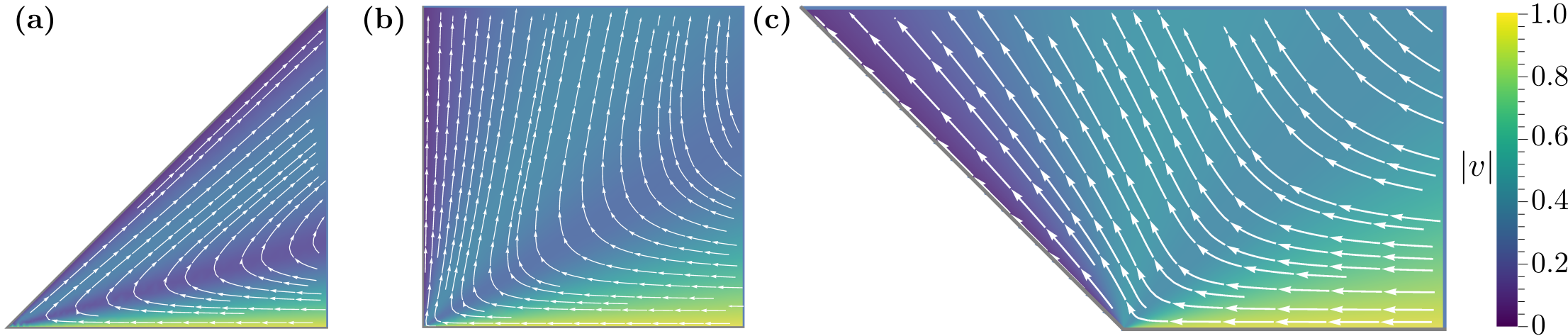}
\caption{{Diffusioosmotic} flow induced by the activity of one phoretic wall in a wedge of angle $\alpha$ for (left to right) $\alpha =\{\pi/4,\pi/2,3\pi/4\}$. Flow streamlines are marked in white. The colour map indicates the total velocity magnitude. {The emergent bulk flow remains comparable in magnitude to the driving slip flow on the active boundary, and decays rapidly close to the inert, no-slip wall.} \label{fig:one_wall}}
\end{figure}

The resulting flow fields are drawn in figure \ref{fig:one_wall} for three values of the wedge opening angle $\alpha =\{\pi/4,\pi/2,3\pi/4\}$. The flow is asymmetric, reflecting the boundary conditions. {We observe the strongest flow along the active boundary, and a gradual decay of the velocity field closer to the other, no-slip wall. We note that the magnitude of the bulk velocity field remains a large fraction of the surface slip velocity, confirming the pronounced effect of surface actuation.}

\subsection{Two phoretic walls}

A simple generalisation of the problem of one wall involves covering both walls of the wedge with catalyst, thereby imposing a slip velocity symmetrically for both $\theta = 0$ and $\theta=\alpha$. In this case, we expect the flow to be symmetrical about the wedge bisector.

The concentration profile satisfying the fixed-flux boundary conditions at both walls, namely,
\begin{subequations}
\begin{align}
    \frac{\partial c}{\partial \theta}=-A\rho \quad &\textrm{for} \quad \theta=0 \\
    \frac{\partial c}{\partial \theta}=A\rho \quad &\textrm{for} \quad \theta=\alpha,
\end{align}
\end{subequations}
follows directly as
\begin{equation}
	c(\rho,\theta) = A \rho \frac{\sin(\theta-\alpha)- \sin\theta}{1-\cos\alpha}.
\end{equation}

Equation \eqref{scraper_ode} remains valid with the new boundary conditions {$F(0)=F(\alpha)=0$ {(no normal velocity at the surface)} and $F'(0)=F'(\alpha)=1$} {(two active walls with non-zero slip velocity)}. The stream function in this case takes on the form
\begin{equation}
	\Psi = \frac{U \rho}{\alpha-\sin\alpha}\left[ (\alpha - \theta)\sin\theta -\theta\sin (\alpha-\theta) \right].
\end{equation}

\begin{figure}
\centering
\includegraphics[width=\textwidth]{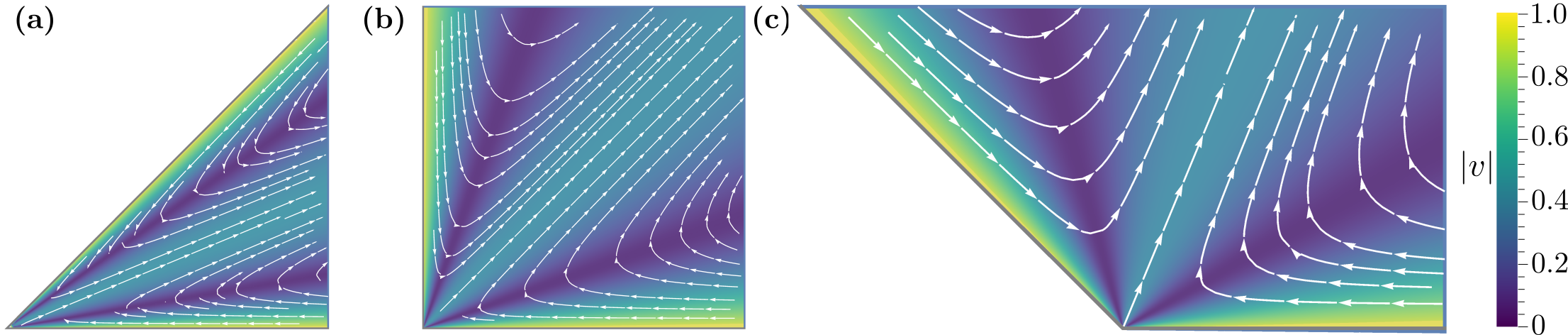}
\caption{{Diffusioosmotic} flow induced by the activity of two phoretic walls in a wedge of angle $\alpha =\{\pi/4,\pi/2,3\pi/4\}$. {The colour map indicates the total velocity magnitude. We note a strong flow close to the driving active boundaries, and a counterflow along the wedge bisector.}
\label{fig:two_walls}}
\end{figure}

The resulting flow fields are drawn in figure \ref{fig:two_walls} for three values of the wedge opening angle $\alpha =\{\pi/4,\pi/2,3\pi/4\}$. The flow is symmetrical, {with the fluid dragged along the walls towards the wedge vertex and expelled along the bisector.}
\section{The diffusion problem with an active and absorbing wall}\label{sec:active-absorbing}

To explore the flow in a more general case, we focus on a corner with a single active sector of length $\ell=b-a$ placed on one of the boundaries at a distance $a$ (between $\rho=a$ and $\rho=b$), as depicted in figure \ref{fig:scheme}. {The other wall is perfectly absorbing.}  Contrary to the previous cases, the problem now has two natural length scales: $\ell$ and $a$. 

The concentration field satisfies {the Laplace equation}, $\nabla^2 c = 0$, with the boundary conditions 
{
\begin{subequations}
\begin{align}
    \frac{\partial c}{\partial \theta} =-A\rho \mathds{1}_{[a,b]} \quad &\textrm{for} \quad \theta=0, \label{condition_theta}\\
    c =0 \hspace{1.5cm} &\textrm{for} \quad \theta =\alpha,
\end{align}
\end{subequations}
}
{where $\mathds{1}_{[a,b]}$ is the indicator function of set $[a,b]$.}
\subsection{The case of $a\neq0$ and $b<\infty$}
The solution in Mellin space is
{
\begin{equation}
    \bar{c}(p,\theta)=A\frac{b^{p+1}-a^{p+1}}{p(p+1)} \frac{\sin(p(\alpha-\theta))}{\cos p\alpha}.
\end{equation}
}
It is assumed that $0<\alpha<2\upi$. The solution has singularities at $p'_k$ such that $p'_k\alpha={\upi}/{2}+k\upi$ for $k\in \mathbb{Z}$, and also at $p_0=-1$, and a removable singularity at $p=0$.  

\begin{figure}
\centering
\hspace{0.05\columnwidth}
\includegraphics[width=0.6\textwidth]{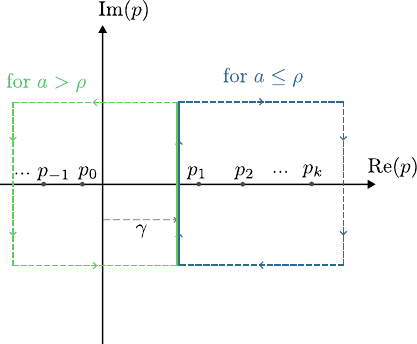}
\caption{{Contours of integration for the evaluation of the inverse Mellin transform. The green contour is used for $a>\rho$, and the blue contour for $a\leq \rho$. Both integration contours are shifted by $\gamma$ along the real axis, and the poles of the integrand are denoted by $p_k$, where $k\in \mathbb{Z}$. To evaluate the integral, one takes the limit of the square side length approaching infinity. In the limit, contributions from the three dashed sides of each square contour vanish, and the desired integral along the imaginary axis can be evaluated using the method of residues.}
\label{fig:contour_diagram}}
\end{figure}

{To evaluate the concentration field, we must now invert the Mellin transform.} The concentration {in the real space can be} written as $c=I_b-I_a$, where
{
\begin{align}
    f_a &=A\frac{a}{p(p+1)} \frac{\sin(p(\alpha-\theta))}{\cos p\alpha} \left(\frac{a}{\rho}\right)^p,  \\ 
    I_a &=\frac{1}{2\upi i}\int_{\gamma-i\infty}^{\gamma+i\infty} f_a dp. \label{eq:gamma}
\end{align}
}
{
The {complex integral above} can be evaluated by applying Cauchy's residue theorem to a square integration contour sketched in figure~\ref{fig:contour_diagram}. For $a \leq \rho$, the integration contour lies in the half-plane $\Real(p) > \gamma$. For $a>\rho$, the appropriate contour lies in the $\Real(p) < \gamma$ half-plane. Such a choice ensures that the integral over three sides of the square contour vanishes when the contour is extended to infinity.}

{In particular, for the integral in Eq.~\eqref{eq:gamma}, we select $\gamma=0$.} The poles are listed as 
\begin{equation}
  p_k= 
    \begin{cases}
    \displaystyle  {\alpha^{-1}\left(\frac{\upi}{2}+(k-1)\upi\right)} & \text{for $k\in \mathbb{Z}_+$}\\
         -1 & \text{for $k=0$}\\
    \displaystyle   \alpha^{-1}{\left(-\frac{\upi}{2}+(k+1)\upi\right)}       & \text{for $k\in \mathbb{Z}_-$}.
    \end{cases}      
\end{equation}
Under the assumption that, for all $k$, $({\frac{\upi}{2}+k\upi})/{\alpha}\neq-1$ (i.e. $\alpha\neq {\upi}/{2},\upi, {3\upi}/{2}$) the residues are
\begin{subequations}
\begin{align}
      \textrm{for }& k\neq0 : \quad \mathop{\mathrm{Res}}_{p = p_k} f_a= -\frac{Aa}{\alpha p_k(p_k +1)}\cos(p_k\theta) \left(\frac{a}{\rho}\right) ^{p_k},\\
      \textrm{for }& k=0 : \quad \mathop{\mathrm{Res}}_{p = p_k} f_a=A \frac{\sin(\alpha-\theta)}{\cos\alpha} \rho.
 \end{align}
\end{subequations}
Depending on the ratio of ${a}/{\rho}$, different residues are contained in the integration contour, and the orientation of the contour is different, hence
\begin{equation}
  I_a(\rho,\theta)= 
    \begin{cases}
     \phantom{-} \sum_{k\leq0} \mathop{\mathrm{Res}}_{p = p_k} f_a & \text{for $a>\rho$}\\
         -\sum_{k\geq1} \mathop{\mathrm{Res}}_{p = p_k} f_a & \text{otherwise,}
    \end{cases}     
\end{equation}
with the full solution being
\begin{equation}
    c(\rho,\theta)=I_b(\rho,\theta)-I_a(\rho,\theta).
\end{equation}
Note that the residue at $p=-1$ is responsible for satisfying the condition \eqref{condition_theta}.
In the special cases of $\alpha= {\upi}/{2},\upi, {3\upi}/{2}$ solutions can be obtained by taking, for example, the limit of $\alpha \to {\upi}/{2}$.
The solutions for $\alpha={\upi}/{2}-\epsilon$ and $\alpha={\upi}/{2}+\epsilon$ can then be compared to check if the desired accuracy has been achieved for a chosen {small parameter} $\epsilon$.

\subsection{ The case of $a=0$, $b<\infty$ i.e. sector $[0,b]$}

The solution is obtained similarly to the previous case as
\begin{equation}
  c(\rho,\theta)= 
    \begin{cases}
     \phantom{-} \sum_{k\leq0} \mathop{\mathrm{Res}}_{p = p_k} f_b & \text{for $b>\rho$},\\
         -\sum_ {k\geq1} \mathop{\mathrm{Res}}_{p = p_k} f_b & \text{otherwise.}
    \end{cases}      
\end{equation}

\subsection{The case of $a\neq0$ and $b=\infty$ i.e. sector $[a,\infty[$ }
In this case existence of Mellin transform requires that ${\gamma} {<}-1$.\\
We thus take any {$\gamma$ such that $-1 > \gamma> -{{\upi}/{2\alpha}}$} and write the solution as
\begin{equation}
  c(\rho,\theta)= 
    \begin{cases}
      -\sum_{k\leq-1} \mathop{\mathrm{Res}}_{p = p_k} f_a & \text{for $a>\rho$},\\
    \phantom{-}  \sum_{k\geq0} \mathop{\mathrm{Res}}_{p = p_k} f_a & \text{otherwise.}
    \end{cases}      
\end{equation}
A simple computation shows that, in all the cases above, $c(\rho, \theta)$ with an appropriate choice of $\gamma$, satisfies the condition for the existence of the inverse transform.

\begin{figure}
\centering
\includegraphics[width=\textwidth]{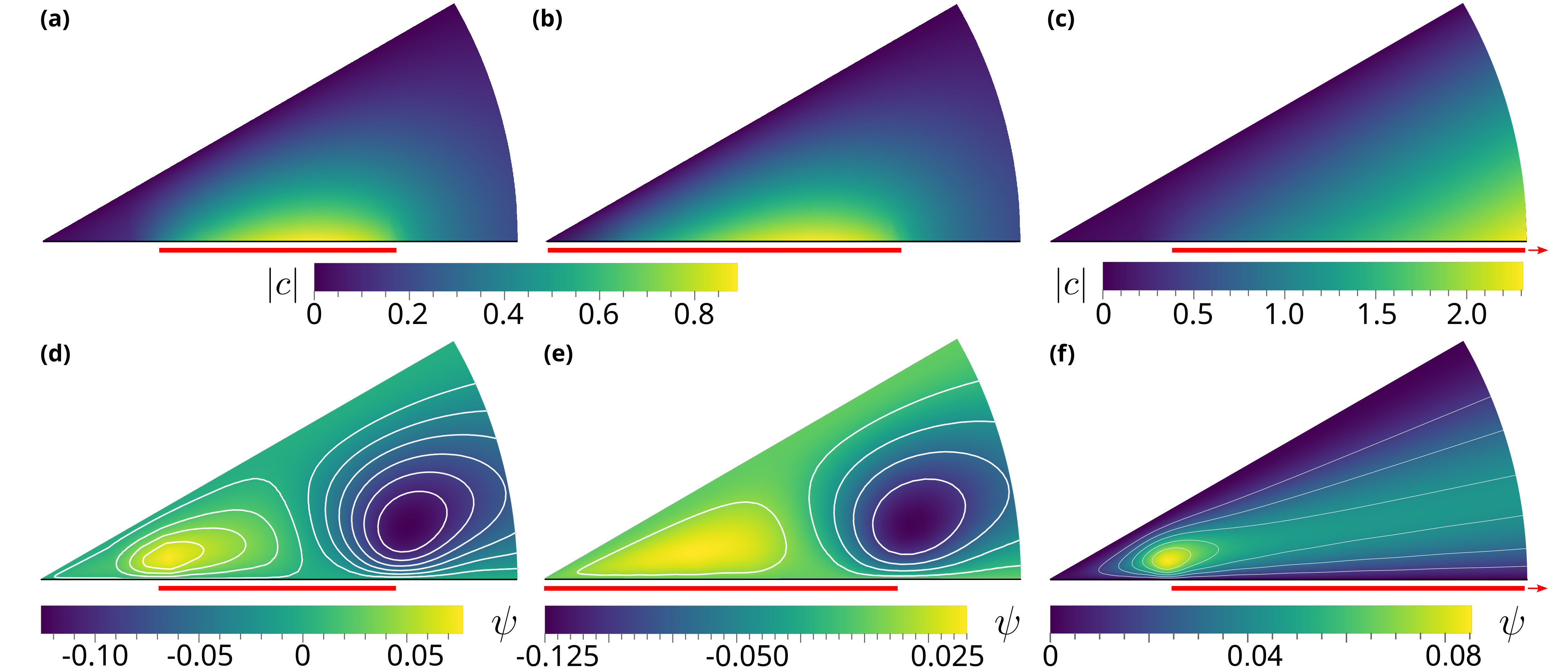}
\caption{{Solute concentration fields, (a)--(c), and isolines of the stream function $\psi$, (d)--(f),} for an ideally absorptive wall at $\theta=\upi/6$, a {catalytic (active)} wall at $\theta=0$ with a catalytic sector at $(a,b)= \{ (1,3), (0,3), ({1},\infty)\}$ marked in red. Here, we assumed $A=1$ and the plotted radius of the wedge is $\rho<4$. The scale bar for the absolute concentration field, $|c|$, is common for plots (a), (b) and different for panel (c).  \label{fig:absorptive_wall}}
\end{figure}

In figure \ref{fig:absorptive_wall}, we sketch the solutions for the concentration field for a chosen wedge opening angle $\theta=\upi/6$ in three representative cases. The catalytic sector is marked in red, and the colours indicate the absolute value of solute concentration. In all cases, the resulting concentration field is heterogeneous, which offers the possibility to drive slip flow along the active boundary.

\section{The diffusion problem with an active and a reflective wall}\label{sec:active-reflective}

\subsection{Solution for a single catalytic sector with reflective walls}
Consider a setting in which the wall at $\theta= 0$ contains a catalytic sector with activity $A$ and the wall at $\theta=\alpha$ is not active, thus it is a reflective wall, with no-flux boundary condition. We denote the corresponding solute concentration field by $c_0$. The boundary conditions are then
\begin{subequations}
\begin{align}
    \frac{\partial c_0}{\partial \theta}=-A\rho \mathds{1}_{{[a,b]}} \quad &\textrm{for} \quad \theta=0  \\
    \frac{\partial c_0}{\partial \theta}=0 \quad &\textrm{for} \quad \theta =\alpha.
\end{align}
\end{subequations}
We again assume that $0<\alpha<2\upi$. A solution satisfying these boundary conditions in the Mellin space reads
\begin{equation}
    \bar{c}_0(p,\theta)=-A\frac{{b^{p+1}-a^{p+1}}}{p(p+1)} \frac{\cos(p(\alpha-\theta))}{\sin(p\alpha)}. \label{eq:catalytic-single}
\end{equation}

If $c_\alpha$ is a solution for the active sector with activity $A$ on the wall $\theta=\alpha$ i.e., for the boundary conditions 
\begin{subequations}
\begin{align}
    \frac{\partial c_\alpha}{\partial \theta}=0 \quad &\textrm{for} \quad \theta=0  \\
    \frac{\partial c_\alpha}{\partial \theta}=A\rho \mathds{1}_{{[a,b]}} \quad &\textrm{for} \quad \theta =\alpha,
\end{align}
\end{subequations}
then the corresponding solution takes the form
\begin{equation}
\quad \bar{c}_\alpha(p,\theta)=\bar{c}_0 (p,\alpha-\theta)=-A\frac{{b^{p+1}-a^{p+1}}}{p(p+1)} \frac{\cos(p\theta)}{\sin(p\alpha)}.
\end{equation}

Let us introduce the notation
\begin{align}
    g_{{a}}(p,\theta)&=-A\frac{a}{p(p+1)} \frac{\cos(p(\alpha-\theta))}{\sin(p\alpha)}\left(\frac{a}{\rho}\right)^p, \\
    I_{{a}}&=\frac{1}{2\upi i}\int_{\gamma-i\infty}^{\gamma+i\infty} g_{{a}} dp.
\end{align}
Applying the inverse Mellin transform to the solution \eqref{eq:catalytic-single}, we can write 
\begin{equation}
    c_0=I_{{b}}-I_{{a}}=\frac{1}{2\upi i}\int_{\gamma-i\infty}^{\gamma+i\infty} (g_{{b}}-g_{{a}}) dp. 
    \label{definition_c_0}
\end{equation}
We now note that $g_{{a}}$ has poles at $p'_k={k\upi}/{\alpha}$, with $k\in \mathbb{Z}$, and at $p=-1$. 
Under the assumption that for all $k$: $p'_k\neq-1$ (that is, $\alpha\neq \upi$) the residues are
\begin{subequations}
\begin{align}
      \textrm{for }& k\neq0 : \quad \mathop{\mathrm{Res}}_{p = p_k} g_{a}= -\frac{Aa}{\alpha} \frac{\cos(p_k(\alpha-\theta))}{p_k(p_k +1)}\left(\frac{a}{\rho}\right) ^{p_k},\\
      \textrm{for }& k=0 : \quad \mathop{\mathrm{Res}}_{p = {0}} g_a= -\frac{Aa}{\alpha}\left[\log\left(\frac{{a}}{\rho}\right)-1\right],\\
      \textrm{for }& p=-1 : \quad \mathop{\mathrm{Res}}_{p = -1}g_a= -A\rho \frac{\cos(\alpha-\theta)}{\sin\alpha}.
 \end{align}
\end{subequations}
The choice of $\gamma$ such that {$\textrm{max}\{p'_k: p'_k<0\}=-{\upi}/{\alpha}<\gamma<0$, $\gamma>-1$}, guarantees that the Mellin transform of $c_0$ exists. Integrating over a square contour in one of the half-planes, we obtain
\begin{equation}
  I_{{a}}(\rho,\theta)= 
    \begin{cases}
      \sum_{k\leq {-1}} \mathop{\mathrm{Res}}_{p = p'_k} g_a +{\mathrm{Res}_{p = -1}g_a} & \text{for $a>\rho$},\\
         -\sum_{k\geq {0}} \mathop{\mathrm{Res}}_{p = p'_k}g_a & \text{otherwise.}
    \end{cases}      
\end{equation}
Using Eq. \eqref{definition_c_0}, $c_0$ can now be computed straightforwardly.
The total activity of the walls not being zero does not cause a contradiction, since there are no constraints on activity at $\rho=\infty$ and the solute can be absorbed or emitted at infinity. The case $\alpha= \pi$ can be solved in Cartesian coordinates or by taking the limit of $\alpha \to \upi$ in the above solution.

\subsection{Diffusion for multiple catalytic sectors}
By linearity, the solution can be obtained by superposition of solutions for single sectors. If sectors $[a_i, b_i]$ have activity $A_i$, the condition of equilibrium is 
\begin{equation}
   \sum_i A_i (b_i-a_i)=0,
   \label{cond1}
\end{equation}
where the sum is taken over all active sectors.
This is also the condition for the solution to be physical (solute conservation). That is, unless we allow the solute to be released/absorbed at $\rho=\infty$. It can be seen that the resulting solutions are indeed continuous. Note that the sectors can overlap, so step activity profiles can be created to mimic arbitrary activity functions $A_0 (\rho)$ and $A_\alpha (\rho)$ of the walls. 

\begin{figure}
    \centering
    \includegraphics[width=\textwidth]{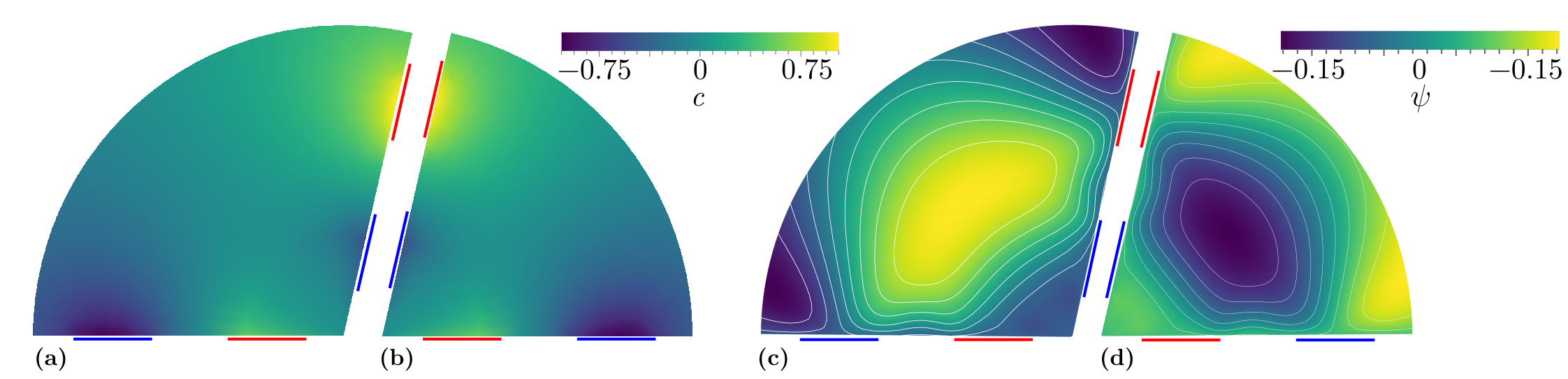}
    \caption{{Solute concentration fields, (a)-(b), and corresponding stream functions $\psi$, (c)-(d), for wedges with multiple active sectors. The wedge angles are $\alpha= \{4\upi/7,3\upi/7 \}$. There are two active sectors of opposite activity $|A|=1$ on each wall. They span the sections $\rho\in(0.5,1.5)$ and $\rho\in(2.5,3.5)$. The plotted radius of the wedge is $\rho<4$. Patches of positive activity are marked in red, while those of negative activity are marked in blue. Panels (a)-(b) share the concentration scale bar placed in the middle of the figure, while the stream function scale bar for panels (c)-(d) is in the top right corner.}}
    \label{fig:multiple_sectors}
\end{figure}

An example of a setting involving multiple sectors is presented in figure \ref{fig:multiple_sectors} for two active patches of opposite activity on each of the walls, and for different wedge opening angles. The effect of activity of patches closer to the tip of the wedge is diminished by their mutual influence, and the concentration fields produced by patches further away from the confinement are more pronounced. This demonstrates the intricate interaction between the geometry of the wedge and the chemical activity of its walls.

\subsection{The diffusion problem for an arbitrary catalytic wall}

Let us now consider a setting where the wall at $\theta= 0$ has arbitrary activity $A=A(\rho)$ and the wall at $\theta=\alpha$ is reflective. 
We will approach this problem using the Green's function {method~\citep{Arfken2013_Green's_functions}}, because the Stokes equation is linear. {Let $c^G_0 (\rho, \rho', \theta)$ be the Green's function, which is a formal solution to the Laplace equation with a point activity $\delta (\rho - \rho')$ at $\rho = \rho'$, $\theta=0$.} Boundary conditions {for $c^G_0 (\rho, \rho', \theta)$} take the following form,
\begin{subequations}
\begin{align}
    \frac{\partial c^G_0}{\partial \theta}(\rho, \rho', \theta)=-\rho \delta(\rho-\rho') \quad \textrm{for} \quad \theta=0  \\
    \frac{\partial c^G_0}{\partial \theta}(\rho, \rho', \theta)=0 \quad \textrm{for} \quad \theta =\alpha.
\end{align}
\end{subequations}
The solution in the Mellin space reads
\begin{equation}
    \bar{c}^G_0(p,\rho',\theta)=-\frac{\cos(p(\alpha-\theta))}{p \sin(p\alpha)} \rho'^p.
\end{equation}
{$\bar{c}_0 (p,\theta)$ is obtained by accounting for all point activities $A(\rho')\delta(\rho-\rho')$ along the boundary}
\begin{equation}
    \bar{c}_0 (p,\theta)=\int_0^\infty A(\rho') \bar{c}^G_0(p,\rho',\theta) d\rho'.
\end{equation}
Residues of $\bar{c}^G_0$ are $p_k={k\upi}/{\alpha}$. We now change the order of integration in the back-transformed concentration field to arrive at
\begin{equation}
c_0 (\rho,\theta)=\frac{1}{2\upi i}\int_0^\infty d\rho' A(\rho') \int_{\gamma-i\infty}^{\gamma+i\infty} -\frac{\cos(p(\alpha-\theta))}{p \sin(p\alpha)} \left( \frac{\rho'}{\rho} \right)^p dp,
\end{equation}
where $\gamma$ has to satisfy the same condition as in the previous section. Defining 
\begin{equation}
    {h}(p,\rho,\rho',\theta)= -\frac{\cos(p(\alpha-\theta))}{p \sin(p\alpha)} \left( \frac{\rho'}{\rho} \right)^p,
\end{equation}
we evaluate the following residues
\begin{equation}
  \mathrm{Res}_{p=p_k} {h} = 
    \begin{cases}
      -\frac{1}{\alpha}\frac{\cos (p_k(\alpha-\theta))}{p_k}\left( \frac{\rho'}{\rho} \right)^{{p_k}} & \text{for $k\neq 0$},\\
      -\frac{1}{\alpha} \log\left( \frac{\rho'}{\rho} \right) & \text{for $k=0$}.
    \end{cases}      
\end{equation}
Choosing the contours of integration the same way as in the previous section, we finally obtain
\begin{equation}
c_0 (\rho,\theta)= \int_{{\rho}}^{{\infty}} d\rho' A(\rho') \sum_{k\leq {-1}} \mathrm{Res}_{p_k}{h}(p,\rho,\rho',\theta)  - \int_{{0}}^{{\rho}} d\rho' A(\rho') \sum_{k {\geq}0} \mathrm{Res}_{p_k}{h}(p,\rho,\rho',\theta).
\end{equation}
We note that for two arbitrary catalytic walls, superposition of solutions can be used. The condition to satisfy solute conservation is
\begin{equation}
    \int_0^\infty (A_0(\rho) + A_\alpha (\rho))d\rho =0, \label{eq:solute_conservation}
\end{equation}
where $A_0 (\rho)$ and $A_\alpha (\rho)$ are the activities of the walls $\theta=0$ and $\theta=\alpha$, respectively. It is worth noting that, for practical purposes, approximating the solution, for a given function $A(\rho)$, by making a superposition of solutions for constant activity sectors could potentially be more efficient than truncating the above series and evaluating the integrals.

\section{Solution of the biharmonic equation for given solute concentration field} \label{sec:biharmonic}
\begin{figure}
    \centering
    \includegraphics[width=\columnwidth]{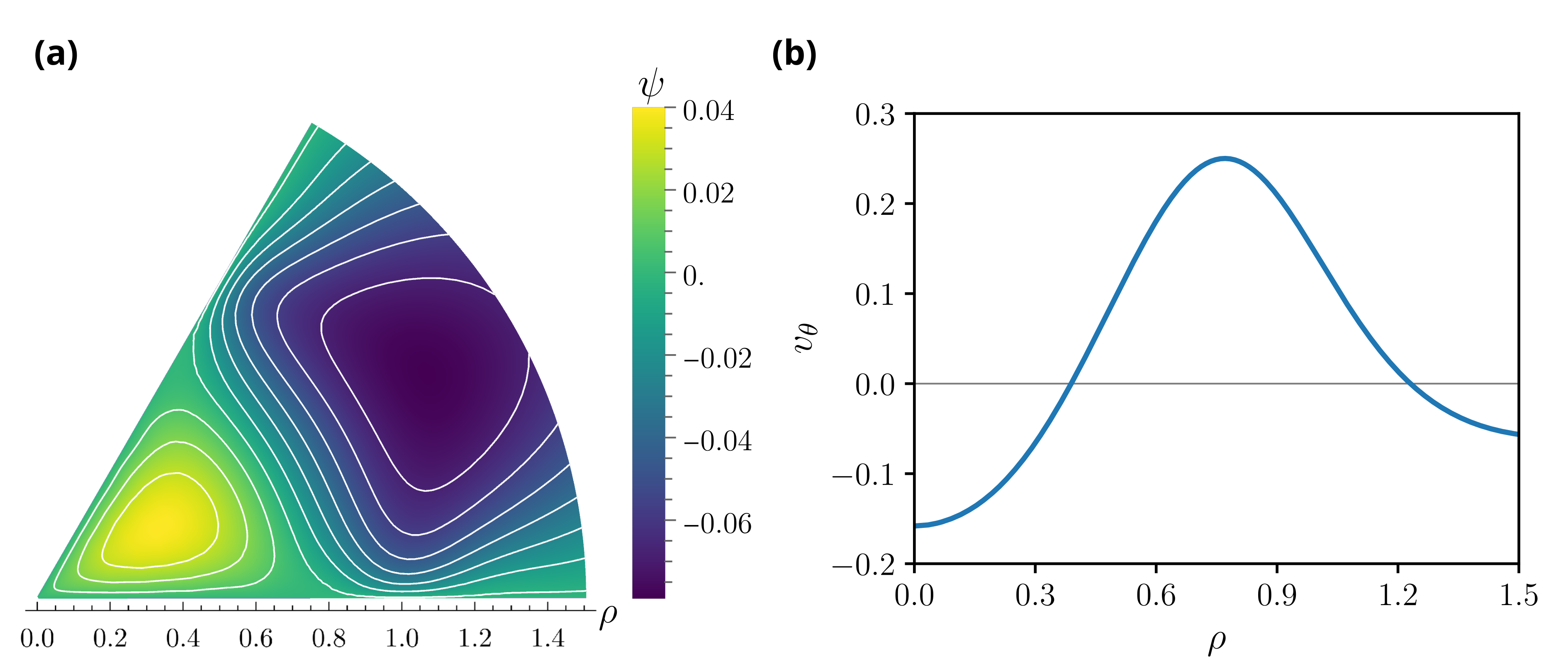}
    \caption{{\bf (a)} Isolines of the stream function $\psi$ of {diffusioosmotic} corner flow for $\theta=\upi/3$, with active sectors at $\rho \in (0,1)$, emitting at $\theta = 0$ and absorbing at $\theta = \alpha$.  The flow in the corner eddy (yellow) is counter-clockwise and drives clockwise rotation of another eddy further away from the corner. Colours code the magnitude of $\psi$. {\bf (b)} Transversal velocity profile $v_\theta(\rho)$, on the bisector angle of the wedge. Roots of the velocity indicate the centres of vortices.}
    \label{fig:stream_function}
\end{figure}

Having resolved the concentration field in various cases and for different boundary conditions on each wall, we now turn to the calculation of the corresponding flow field. The boundary conditions for the stream function $\Psi(\rho,\theta)$, {which} satisfies the biharmonic equation \eqref{eq:biharmonic}, are
\begin{subequations}
    \begin{align}
        u_\theta=-\frac{\partial \Psi}{\partial \rho}=0  \quad \textrm{for} \quad \theta=0,\alpha         \\
        u_\rho=\frac{1}{\rho}\frac{\partial \Psi}{\partial \theta}=M\frac{\partial c}{\partial \rho}  \quad \textrm{for} \quad \theta=0,\alpha,
    \end{align}
\end{subequations}
indicating that the only source of flow is the slip flow at the boundary, dictated by the heterogeneous solute concentration profile. Applying the Mellin transform $\mathscr{M}_p$ to these conditions yields
\begin{subequations}
    \begin{align}
        \bar{\Psi}=0 \quad \textrm{for} \quad \theta=0,\alpha         \\
         \frac{\partial \bar{\Psi}}{\partial \theta}=-pM\bar{c} \quad \textrm{for} \quad \theta=0,\alpha.
    \end{align}
\end{subequations}
Using the general form of the solution for $\bar{\Psi}$, given by Eq.~\eqref{psi_general}, we may now identify the coefficients to be
\begin{subequations}
\begin{align}
    G_1&=-F_1 \hspace{210pt} &\\
         \begin{bmatrix}
            F_1\\
            F_2\\
            G_2
        \end{bmatrix}
    &=\frac{-p M}{2D(p)}
    \bar{c}(p,0)
        \begin{bmatrix}
            (p+1) \sin(2\alpha) - \sin(2 \alpha (p+1))\\
            (p+1) \cos (2 a)+\cos (2 a (p+1))-p-2\\
            -p+(p+1) \cos(2\alpha) - \cos(2\alpha(p+1))
        \end{bmatrix}
    +\notag
    \\
    &\phantom{=} +\frac{-p M}{2D(p)}
    \bar{c}(p,\alpha)
        \begin{bmatrix}
             (p+2) \sin (\alpha p)-p \sin (\alpha (p+2)) \\
             -((p+2) (\cos (\alpha p)-\cos (\alpha (p+2)))) \\
             p (\cos (\alpha p)-\cos (\alpha (p+2))) 
        \end{bmatrix}
     ,
\end{align}
\end{subequations}
where $D(p)=(p+1)^2 \cos (2\alpha) -p(p+2) -\cos(2\alpha(p+1))$. The way to obtain $\bar{c}(p,0)$ and $\bar{c}(p,\alpha)$ for an arbitrary setting of catalytic sectors on both walls has been demonstrated in previous sections. The remaining task is thus to invert the Mellin transform; this turns out to be challenging analytically, since zeros of $D(p)$ are difficult to obtain, unlike when inverting the Mellin transform for the concentration. 
We thus compute the integral (see Eq.~\eqref{eq:reverse_Mellin}) numerically. In an illustrative case, where the active sectors on the walls span from the corner to $\rho=1$, the calculated streamlines (the isolines of $\Psi(\rho,\theta)$) can be seen in figure \ref{fig:stream_function}. A counterclockwise eddy is created in the immediate vicinity of the active patches, and its rotation drives a larger eddy in the opposite direction further away from the corner. We note that this flow profile is obtained using only two active patches. When combined into more complicated coverage patterns, the flow can be tuned to a particular application, e.g. localised mixing.

\begin{figure}
    \centering
    \includegraphics[width=\columnwidth]{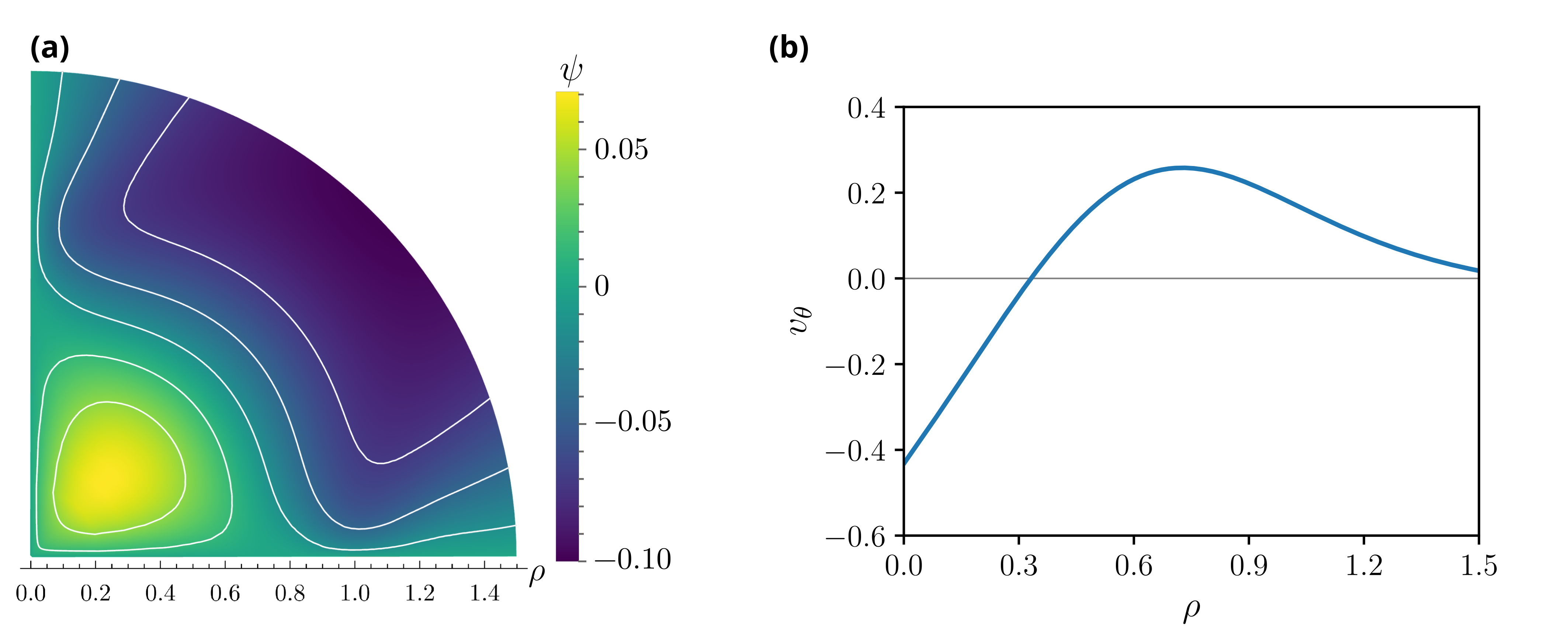}
    \caption{{Diffusioosmotic corner flow for the wedge opening angle of $\upi/2$. {\bf (a)} Isolines of the stream function show a single vortical structure in the corner. The active patches {cover} both walls close to the tip (for $\rho<1$), with emission at the horizontal wall and absorption at the vertical surface. {\bf (b)} Transversal velocity profile along the bisector line of $\theta=\upi/4$.}   }
    \label{fig:stream_function_90}
\end{figure}

{It is useful to consider an experimentally practical scenario of the wedge opening angle of $\alpha=\upi/2$, for which we present the flow field in figure \ref{fig:stream_function_90}. In this case, a single vortical structure emerges, fueled by catalytic reactions on the walls close to the corner. The active patches cover the area of $\rho<1$ on both walls. The associated transversal velocity profile confirms that the vortex does not extend far beyond the active region.} 

\section{{Diffusion} with analytic Neumann boundary conditions} \label{sec:neumann}

Here, we present an auxiliary result for the diffusion problem when the activity patterns are given by analytic functions, rather than step-like profiles discussed earlier. 
Let $A{^{(1)}} (\rho)$ and $A{^{(2)}} (\rho)$ represent the (analytic) distribution of activity on the walls with $\theta=0$ and $\theta=\alpha$, respectively. We make the following ansatz for the concentration profile $c$ in the fluid domain
\begin{equation}
    c={\lambda_0} +\sum_{n\geq 1} (F_n \cos(n\theta) +G_n \sin(n\theta))\rho ^n, \label{eq:expansion}
\end{equation}
and expand $A^{(1)}$ and $A^{(2)}$ in power series in $\rho$ as
\begin{subequations}
    \begin{align}
        A^{(1)}=\sum_{n\geq 1} A^{(1)}_n \rho^n,\
        A^{(2)}=\sum_{n\geq 1} A^{(2)}_n \rho^n.
    \end{align}
\end{subequations}
Imposing the fixed flux boundary conditions, Eqs. \eqref{condition_0} and \eqref{condition_alpha}, we arrive at linear equations for the coefficients of the expansion in Eq.~\eqref{eq:expansion}. Solving these equations, we obtain
\begin{subequations}
    \begin{align}
        F_n&=-\frac{A^{(1)}_{n-1}\cos(n\alpha)+A^{(2)}_{n-1}}{n\sin(n\alpha)},\\
        G_n&= -\frac{A^{(1)}_{n-1}}{n},
    \end{align}
\end{subequations}
for $n\geq 1$. While {$\lambda_0$} remains arbitrary, it can be fixed by an additional boundary condition, such as the value of $c(\rho=0)$. Then, we get $c(\rho=0)={\lambda_0}$. The condition for existence of $F_n$ and $G_n$ such that boundary conditions can be satisfied by this ansatz is $\sin(n \alpha) \neq 0$ (i.e. $\alpha \neq q \upi$, where $q \in \mathbb{Q}^+$). If $\sin(n \alpha) = 0$, a more general ansatz is required. Alternatively, we note that the above ansatz is sufficient if $\alpha=r \upi$ for $r\in \mathbb{R}^+\backslash \mathbb{Q}^+$. Then $r$ can taken to be, for example, $r=q \pm \sqrt{2}\cdot 10^{-n}$, where $n$ can be arbitrarily large; Then we obtain a solution for the desired angle $\alpha=q \upi$ in the limit of large $n$.

\section{Discussion and conclusions} \label{sec:conclusions}
In this paper, inspired by Moffatt eddies~\citep{Moffatt_1964_JFM_18_1, Moffatt_1964_AMS_2_365} {that emerge in wedge-like geometries} under {mechanical} boundary forcing, we have shown that corner geometry with chemically active sectors can be used to generate similar eddies. Our solution involved the exact solution of the diffusion problem in the wedge geometry using the Mellin transform formalism, followed by a solution of the biharmonic equation for the stream function of the associated flow. The proposed {diffusioosmotic framework relies on a fixed-flux boundary condition for the solute release from a catalytic patch, which is appropriate when the diffusion of the solute is much faster than its adsorption (that is, in the limit of low Damköhler number) and that the advection is slow compared to diffusion (corresponding to the limit of low P\'eclet number).}

In the simple case of uniform coverage of the walls with chemical activity, the geometric asymmetry alone is responsible for heterogeneous distribution of solute concentration, which drives the flow. In such a setting, the problem admits a simple analytical solution akin to the classical Taylor's scraper problem when one wall is chemically active. Similarly, an analytical solution is also available when both walls are covered with catalyst. These solutions can serve as benchmarks for numerical computations, particularly when solving for the concentration field with non-trivial {active} boundary conditions.

In the more general case, where patches of activity are distributed on the walls, we presented a solution technique using the Mellin transform to solve the diffusion problem and,  subsequently, find the coefficients of the general solution of the associated flow problem. The solutions can be evaluated numerically for an arbitrary distribution of patches and a combination of boundary conditions. 

{We note here that while our present work bears resemblance to the previously analysed case of flow in a 2D confined channel of \citet{Visan2024} in the limit of $\alpha\to\upi$ with a single active sector, a straightforward comparison is not possible. In our case, there is a single wall, and the concentration field decays monotonically with the distance from the active patch, while \citet{Visan2024} consider a system confined between two walls; the bottom surface is patterned with activity, and they impose a Dirichlet boundary condition of constant concentration on the top wall. Additionally, they assume periodic boundary conditions in the direction parallel to the walls. They then observe a pair of counter-rotating vortices. In our system, with only one surface being present for $\alpha=\pi$, we see similar flow structure in the vicinity of an active patch. However, the presence of the top boundary in their system introduces an additional length scale (apart from the size of the active patch itself) that sets the size of recirculating vortices. This interplay of length scales is further discussed by \citet{Michelin2019}. In the absence of the top confining wall, we see different asymptotics of the concentration field (decay with the distance), and the recirculation zone becomes larger but the flow strength decays faster away from the patch.}

{To assess the practical relevance of our theoretical results, we estimate the expected flow velocities in a potential experimental realization of our model. The typical slip velocity, defined in Eq.~\eqref{slip_def}, is given by the product of the surface mobility $\mathcal{M}$ and the normal gradient of solute concentration. Following the classical analysis of Derjaguin \textit{et al.} (1947; reprinted in~\citet{Derjaguin1993}), the magnitude of the surface slip velocity can be approximated as $u \sim (k_B T / \eta) KL^\ast \nabla_\parallel c$, where $\nabla_\parallel c$ is the tangential concentration gradient along the active surface, $k_B T$ is the thermal energy, $\eta$ is the viscosity, and $K$ and $L^\ast$ are parameters characterizing the solute distribution profile in the boundary layer adjacent to the surface. For uncharged solute molecules, this profile arises from surface interactions such as van der Waals and dipole forces, as well as excluded volume effects. Using the representative value $KL^\ast \approx 6\times 10^{-16}\ \text{cm}^2$ reported by~\citet{anderson1989}, together with a realistic concentration gradient of $0.1\ \text{mol}/\text{cm}^4$ in water at room temperature, we obtain slip velocities of approximately $u_s \sim 2\ \mu\text{m}/\text{s}$. This estimate is comparable to typical flow speeds in microchannels. As illustrated in figures~\ref{fig:stream_function} and~\ref{fig:stream_function_90}, the azimuthal velocity component reaches a substantial fraction of this value, indicating that boundary actuation represents a feasible mechanism for flow control.}

{ We note that our solution can also be applied as an approximation in finite geometries, such as that of a rectangular microfluidic channel with chemically patterned corners. When active sectors are contained in the region of a corner (with $\alpha=\upi/2$) such that $\rho<b_\text{max}$, the concentration field away from the active region asymptotically behaves as
$c(\rho,\theta) \sim \left({b_\text{max}}/{\rho}\right)^2$. Thus, the slip velocity due to chemically active sectors scales as $|\bm{u}_s|\sim \left( {b_\text{max}}/{\rho}\right)^3$. 
Note that for such approximation to apply to finite domains, total solute flux away from the chemically active region has to vanish. 
It can be verified that eq.~\eqref{eq:solute_conservation} is a sufficient condition for the total flux to vanish for $\rho>b_\text{max}$, and the local solute current asymptotically scales as $ \left({b_\text{max}}/{\rho}\right)^3$. In the limit of well-separated chemically active corners, one can superpose the solutions from neighbouring corners, and as a result produce the streamline pattern for a more complex channel shape.}

{Our findings thus may have direct implications for microfluidic applications, particularly in small-scale mixing and catalytic processes. Recent studies by~\citet{Munteanu2020} and~\citet{Popescu2025} demonstrated that patches of glucose oxidase (GOX) imprinted on planar surfaces can generate surface slip flows through enzymatic decomposition of glucose solutions. The resulting chemical gradients produce diffusioosmotic slip velocities on the order of $1$–$10\ \mu\text{m}/\text{s}$, providing a realistic example of active surfaces suitable for the geometries considered here. {Moreover, it has been shown that inhomogeneous distribution of solute in dead-end pores can enhance the diffusiophoretic removal of colloids from confined spaces \cite{Li2025}, and the proposed corner design offers a way to achieve this heterogeneity without moving parts.} Although the present framework is developed for osmotic flows, the underlying mechanism is general and may be extended to other phoretic/osmotic effects such as electrophoresis or thermophoresis. These mechanisms can similarly generate surface gradients that drive macroscopic flow in wedge-shaped cavities or microfluidic channels with slanted walls. Given that diffusioosmosis has already been proposed as a mechanism for nanoscale pumping~\citep{Chanda2022} and for guiding the migration of biomolecules, blood cells, and vesicles into microcavities~\citep{VrhovecHartman2018}, we offer here a new geometric configuration for directed transport. When combined with external control methods—such as intensity-tuned photocatalysis~\citep{Timmerhuis}, enzymatic activity~\citep{Popescu2025}, or light-activated diffusioosmosis~\citep{Muraveva2024}—the proposed geometry offers a promising platform for the coupled hydrodynamic and chemical manipulation of particles in confined corner flows.}

\bigskip 

{\bf Funding} {This research was supported by the National Science Centre of Poland Sonata Bis grant no.
2023/50/E/ST3/00465 to ML.}

{\bf Declaration of interests} {The authors report no conflict of interest.}

{\bf Data availability statement} {The Mathematica notebooks used to compute the results of this study are openly available in Zenodo at  https://doi.org/10.5281/zenodo.17800756, reference number 17800756.}

{\bf Author ORCIDs} {M. Lisicki, https://orcid.org/0000-0002-6976-0281; D. Nowak, https://orcid.org/0009-0001-8334-2860}

\end{document}